\documentstyle[sprocl]{article}

\input{psfig}
\newcommand{\BEQ}{\begin{equation}}    
\newcommand{\BEA}{\begin{eqnarray}}
\newcommand{\EEQ}{\end{equation}}      
\newcommand{\EEA}{\end{eqnarray}}
\newcommand{\eps}{\epsilon}                      
\newcommand{\sig}{\sigma}                        
\newcommand{\ups}{\Upsilon}                      
\newcommand{\rar}{\rightarrow}                   

\begin{document}

\title{ON THE PHASE DIGRAM OF THE $q\rar 1$ EXTENDED POTTS MODEL 
AND LATTICE ANIMAL COLLAPSE \footnote{Invited talk presented at the
Seventh Nankai Workshop on Symmetry, Statiscal Mechanics Models and 
Applications, 12-13 Aug. 1995, Nankai University, Tjanjin, China}}

\author{Malte Henkel}

\address{Laboratoire de Physique du Solide \footnote{Unit\'e de recherche
associ\'ee au CNRS no. 155}, Universit\'e Henri Poincar\'e
Nancy I, B.P. 239, F - 54506 Vand{\oe}uvre l\`es Nancy Cedex, France}


\maketitle\abstracts{
The phase diagram of the two-dimensional extended $q-$states Potts model
is investigated in the $q\rar 1$ limit. This is equivalent to 
studying the phase diagram of a two-dimensional infinite interacting
lattice animal. An exact solution on the Bethe lattice and a 
Migdal-Kadanoff renormalization group calculation 
predict a line of $\theta$ transitions from an extended to
a compact phase in the lattice animal. We compare this with the
phase diagram predicted from previous numerical studies. 
}

\begin{figure}[h]
\psfig{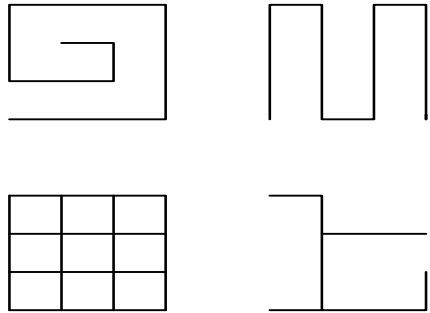}
\caption{Some examples of lattice animals. \label{fig1}}
\end{figure}

The collapse transition of polymers in dilute solution
has been a subject of much current attention.
Here, we are interested in the collapse of interacting, 
two-dimensional branched polymers,
as modelled by lattice animals. A lattice animal is a connected graph 
of occupied sites. Some examples are shown in
Fig.~\ref{fig1}. 
Two neighboring occupied sites may or may not be
immediately connected. If they are, we say there is a {\em bond}
between them. If they are not, we say there is a {\em contact}
between them. 
The statistical mechanics is conveniently
described in terms of a generating function \cite{Fles92,Seno94} 
\BEQ
{\cal Z} = \sum_{n} x^n {\cal Z}_n(y,\tau) = \sum_{n,b,k} a_{n,b,k}
x^n y^b \tau^k \label{GenFun}
\EEQ
where $a_{n,b,k}$ is the number of animals with $n$ sites, $b$ bonds and
$k$ contacts and $x,y,\tau$ are fugacities. 
The infinite lattice animal can be
described either by taking $n\rar\infty$ or by fixing one of the
fugacities in terms of the other two, say $x \equiv x_c (y,\tau)$. 

\begin{figure}[h]
\psfig{figure=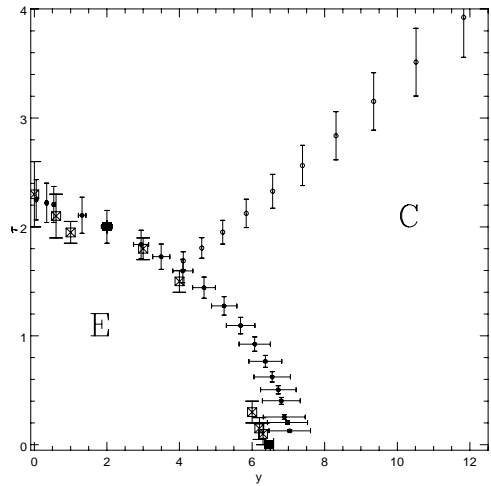,height=7cm,width=8.5cm}
\caption[Phase diagram]{Phase diagram of the infinite 
lattice animal. The dark squares
mark the exactly known percolation critical point and the strong embedding
transition. Open squares denote the results from transfer matrix 
calculation and open and closed circles are the results from graph
enumeration. C labels the compact and E the extended phase. \label{fig2}}
\end{figure}

This system has been extensively studied numerically, using Monte
Carlo simulations \cite{Dick84,Lam87},
exact graph enumerations \cite{Fles92,Chang88,Madr90,Pear95}
and transfer matrix 
methods.\cite{Seno94,Derr83} The results for the phase diagram,
shown in Fig.~\ref{fig2}, 
are however controversial. It is generally agreed upon that, 
as $y$ or $\tau$ is increased, the animal
undergoes a $\theta$-transition from an extended to a compact phase.
The point in question is the possible existence of a phase transition
between two distinct compact phases. 

Physically, compact and
extended phases can be distinguished from the scaling behaviour of
the mean radius of gyration, $\langle R \rangle$, with the number
of monomers $N$, $\langle R \rangle \sim N^{\nu}$. 
In the compact phase,
$\nu=1/d$ and in the extended phase, $\nu=0.6408(3)$ in two 
dimensions.\cite{Derr82} Existing numerical 
results \cite{Fles92,Seno94,Pear95} agree well with
each other on the location and the nature of the extended--compact
transition, see Fig.~\ref{fig2}. At $y=\tau=2$, the $\theta$ point
coincides with the critical point of percolation.\cite{Fles92,Seno94}
In the $y\rar 0$ limit, the strong embedding case is recovered.\cite{Derr83}
This fixed point is unusual for a two-dimensional system as it is not
conformally invariant.\cite{Mill93} 
For $0\leq \tau < 2$, the $\theta$ line should be controlled by the 
strong embedding fixed point,\cite{Seno94,Pear95} which leads to
$\nu=\frac{1}{2}$ and $\phi=\frac{2}{3}$, respectively, where $\phi$ is
the cross-over exponent. On the other hand, the $\theta$ line for $\tau > 2$
should be in a different universality class and it was 
conjectured \cite{Seno94} that $\nu=\phi=\frac{8}{15}$. 

On the other hand, graph enumeration studies \cite{Fles92,Pear95} 
suggest the existence of a further transition between two {\em distinct}
compact phases, a cycle-rich branched polymer for large $y$ and a
contact-rich state for large $\tau$. 
No sign of this transition was found using a transfer
matrix approach.\cite{Seno94} This discrepancy led to the investigations
whose results \cite{Henk95,Henk95a} will be reviewed here.  

We shall approach the problem by using the known \cite{Giri77,HL81,Wu82} 
{\em exact} mapping between the lattice animal and the extended 
$q-$state Potts model described by the classical Hamiltonian
\BEQ \label{Potts}
{\cal H} = - J \sum_{(i,j)} \delta_{\sig_i,\sig_j} 
           - L \sum_{(i,j)} \delta_{\sig_i,1} \delta_{\sig_j,1}
           - H \sum_{i} \delta_{\sig_i,1}
\EEQ
where $\sig_i = 1,2,\ldots,q$ and the fugacities are expressed in terms
of $J,L,H$ as 
\BEQ
x = \exp[ -H - \gamma (J+L) ] \;\; , \;\;
y = \left( e^J -1 \right) e^{J+L} \;\; , \;\;
\tau = e^{J+L} 
\EEQ
where $\gamma$ is the coordination number. 
The relation of the extended 
Potts model (\ref{Potts}) to the lattice animal problem is
\BEQ \label{Corres}
{\cal Z} = 
\lim_{q\rar 1} \frac{\partial}{\partial q} \ln \widetilde{Z} \;\; , \;\;
\widetilde{Z} = \sum_{ \{ \sig \} } e^{-{\cal H} }.
\EEQ 
Thus by calculating the phase diagram of the extended Potts model
information about the phases of the animal problem is obtained. 
The limit $n\rar\infty$ corresponds to finding critical points of 
the Hamiltonian (\ref{Potts}). Transitions between different states of the 
infinite lattice animal 
correspond to multicritical points within the critical manifold
of the extended Potts model.\cite{deGennes79}  

We now investigate the phase diagram of the extended Potts model
(\ref{Potts}) by (a) solving the model exactly on the Bethe lattice and
(b) using the Migdal-Kadanoff renormalization group. The results are
reinterpreted in terms of the lattice animal providing additional
evidence that there is no compact--compact transition.

We first outline the exact solution of the
$q\rar 1$ limit of the extended Potts model (\ref{Potts}) 
on the Bethe lattice.\cite{Henk95,Henk95a} The following
relationship greatly simplifies the calculations: We
are interested in the lattice animal free energy $F = \lim_{n\rar\infty}
n^{-1} \ln {\cal Z}_n$. On the other hand, the animal generating function 
$\cal Z$ diverges at the critical point $x_c$ given by \cite{Seno94}
\BEQ \label{Fxc}
F = - \ln x_c .
\EEQ
Thus, working in the grand canonical ensemble, it is sufficient to find
$x_c$ as a function $x_c(y,\tau$) of the other fugacities $y$ and $\tau$
to obtain the canonical free energy of the {\em infinite} lattice animal. 

Following Baxter,\cite{Baxt82} it is easy to see that on a Bethe lattice
with $\gamma=3$ nearest neighbors for each site and central spin $\sig_0$,
the extended Potts model partition function factorizes into contributions
$Q_n \left( \sig_0 \left| s^{(j)} \right. \right)$
from the j$^{th}$ subbranch with spins $s^{(j)}$ 
\BEQ
\widetilde{Z} = \sum_{\sig_0} e^{H \delta_{\sig_0,1}} 
\sum_{s} \prod_{j=1}^{3} 
Q_n \left( \sig_0 \left| s^{(j)} \right. \right) .
\EEQ
Now the thermodynamics is completely specified 
by $g_n(\sig_0 ) = { \sum_{ \{ s \} } } Q_n \left( \sig_0 | s \right)$
where $n$ refers to the number of iterations performed in the construction
of the Bethe lattice.\cite{Baxt82} Recursion relations for the $g_n$ are
\BEQ \label{GRec}
g_n(\sig_0) = \sum_{s_1=1}^{q} \exp\left( J \delta_{\sig_0,s_1} + 
L \delta_{\sig_0,1}
\delta_{s_1,1} + H \delta_{s_1,1} \right) 
\left( g_{n-1}(s_1) \right)^{2} . 
\EEQ
In analogy with the mean-field treatment of the Potts model,\cite{Bara83} 
the analysis can be simplified by introducing the variables
\BEQ \label{Thermo}
\Xi_n = \frac{ g_n(\sig_0 \neq 1,2,3) }{g_n (1) } \;\; , \;\; 
\ups_n = \frac{ g_n (2) }{ g_n (1) } \;\; , \;\;
Z_n = \frac{ g_n (3) }{ g_n (1) }.
\EEQ
Besides simplification, the introduction of these variables provides
a sufficient number of order parameters to enable the model to have
a phase diagram rich enough to be compared with Fig.~\ref{fig2}. 
As $n\rar\infty$, these variables tend to fixed point values
$\Xi, \ups, Z$, which can be determined from the recursion relations 
(\ref{GRec}). At this
stage, we take the $q\rar 1$ limit (see (\ref{Corres})) and find the
self-consistency relations 
\newpage
\BEA 
\Xi &=& \frac{ x^{-1}\tau^{-3} + \ups^2 + Z^2 + (y/\tau -2) \Xi^2}
{x^{-1}\tau^{-2} + \ups^2 + Z^2 - 2 \Xi^2 } ,\nonumber \\
\ups &=& \frac{ x^{-1}\tau^{-3} + ( y/\tau +1)\ups^2 + Z^2 - 2 \Xi^2}
{x^{-1}\tau^{-2} + \ups^2 + Z^2 - 2 \Xi^2 } , \label{SelfCon} \\ 
Z &=& \frac{ x^{-1}\tau^{-3} + \ups^2 + (y/\tau+1) Z^2 - 2 \Xi^2 }
{x^{-1}\tau^{-2} + \ups^2 + Z^2 - 2 \Xi^2 } .\nonumber
\EEA
These equations can be decoupled and reduced to an equation for the
fugacity $x$ which is at most quadratic.\cite{Henk95a} 
Using eq.~(\ref{Fxc}), we can then read off the free energy. 

Thus we obtain $x=x(p;y,\tau)$ as a
function of $p=(\Xi+\ups)\tau$ which plays the role of an order
parameter. The value of $p$ is fixed by maximising $x(p)$ with respect
to $p$. This gives the critical surface of the extended Potts model. 
A detailed analysis \cite{Henk95a} of eqs.~(\ref{SelfCon}) shows that
the equilibrium phases can be described in terms of two functions
$x=x(p;y,\tau)$. 
One is 
\BEQ
A(p;y,\tau) = x^{(A)} (p;y,\tau) = \frac{2(p-2)}{y p^2}.
\EEQ
and corresponds to the extended phase and the other one is
\BEA
C_{-}(p;y,\tau) &=& x_{}^{(C)}(p;y,\tau) \\
&=& \frac{2(1-\tau)p-y^2/4}{py(p^2-4\tau p-y^2/4)}
- \frac{\sqrt{p(4\tau^2-8\tau+4+y^2/4)-y^2/4}}
{\sqrt{p} (p^2-4 \tau p-y^2/4) y}. \nonumber
\EEA
and corresponds to the compact phase. 
In particular, it can be shown \cite{Henk95a} that if two of the variables
$\Xi,\ups,Z$ are equal, all three of them have to coincide. 
Therefore only two distinct phases are possible for the 
infinite lattice animal. 
For case A
\BEQ
F = -\ln x^{(A)} = - \ln \left( \frac{2(p-2)}{p^2 y} \right) \;\; , \;\;
\frac{\partial F}{\partial p}= \frac{p-4}{p(p-2)} \;\; , \;\;
\left. \frac{\partial^2 F}{\partial p^2}\right|_{p=4} = \frac{1}{8} > 0
\EEQ
and $F$ has a single minimum at $p=4$. Indeed, $F|_{p=4}=\ln (4 y)$  
is concave in $y$. The solutions $A,C_-$ meet at $p=4$. 

Thus two distinct phases of the infinite lattice animal are described
by the two solutions $A(p;y,\tau)$ (extended) 
and $C_{-}(p;y,\tau)$ (compact). 
To obtain the transition lines between the two phases, note that
$\partial C_{-}/\partial p|_{p=4} =0$ if and only if $\tau=2$. Furthermore,
$C_{-}(p;y,2)$ for $p=4$ has
a maximum, turning point or minimum for $y>8$, $y=8$ or 
$y<8$, respectively. Thus, for $y<8$, there is a first-order
transition between the extended and compact phases which 
is given by the conditions
\BEQ
C_{-}(p;y,\tau) = A(4;y,\tau) = \frac{1}{4 y} \;\; , \;\; 
\frac{\partial C_{-}}{\partial p} (p;y,\tau) = 0 \;\; , \;\; 
\frac{\partial^2 C_{-}}{\partial p^2} \leq 0 .
\EEQ
At this transition point, $p$ jumps from its value $p=4$ for $\tau$ small
to a new value $p_c(y,\tau)<4$. 
On the other hand, for $y>8, \tau=2$ there is a second order 
transition as $A$ and $C_-$
merge into each other. The second order line ends at $y=8, \tau=2$
in a tricritical point.

Having found a single compact phase from the Bethe lattice calculations,
we now supplement this with a Migdal-Kadanoff renormalization group study. 
Our eventual result that qualitatively the {\em same} phase diagram results
from {\em both} schemes makes it plausible that our analytical approach is
capable of capturing the relevant physics of the system. 

For $d=1+\eps$ dimensions and rescaling factor $b=2$ the 
Migdal-Kadanoff recursion equations for the
extended Potts Hamiltonian (\ref{Potts}) are \cite{Henk95,Henk95a}
\BEA 
\xi^{'} &=&
\left( 
{{\xi(1+\rho+(q-2)\eta)}\over{1+\xi^2+(q-2)\eta^2}}
\right)^{b^{\eps}} \;\; , \;\; 
\eta^{'} =
\left(
{{\xi^2+2\eta+(q-3)\eta^2}\over{1+\xi^2+(q-2)\eta^2}}
\right)^{b^\eps} \;\; , \;\; 
\nonumber \\
\label{Recursion}
\rho^{'}&=&
\left( 
{{\rho^2+(q-1)\xi^2}\over{1+\xi^2+(q-2)\eta^2}}
\right)^{b^\eps} 
\EEA
where
\BEQ
\xi = \exp(H/2 -J) \;\; , \;\; \eta = \exp(-J) \;\; , \;\;
\rho = \exp(L+H).
\EEQ
Eqs.~(\ref{Recursion}) were obtained by performing a 
one-dimensional decimation followed by bond-moving.\cite{Bur82}

\begin{figure}
\psfig{figure=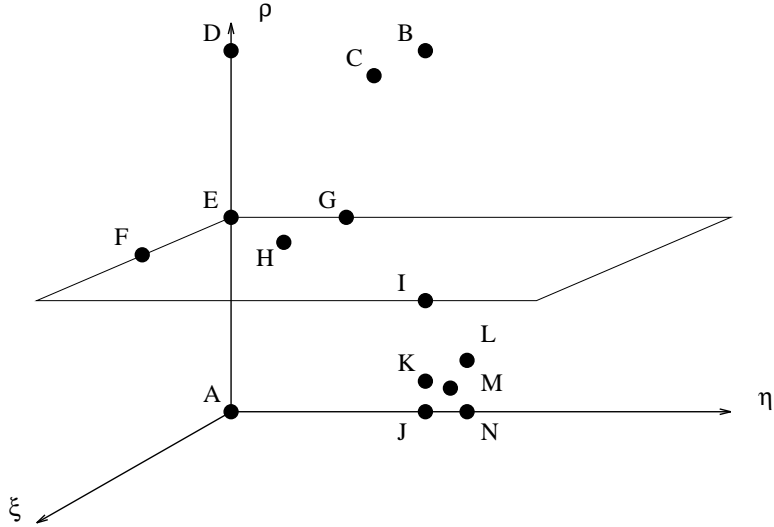,height=7cm}
\caption[MK fixed points]{Fixed points of the MK recursion 
relations. The plane $\rho=1$ is shown to guide the eye. \label{fig3}}
\end{figure}

The fixed point structure that follows \cite{Henk95,Henk95a} 
from the recursion
equations (\ref{Recursion}) is complicated and 
$q$-dependent. However in the two limits of interest to us ($q \to 1$ and
$\eps \to 0$) a clear pattern appears and 14
fixed points can be identified. They are shown in Fig.~\ref{fig3}. 
They are interpreted using existing results \cite{Kauf81} on the
$q$- state Potts model as a guide. 

We point out that the fixed point structure changes {\em at} $q=1$, as can
be seen by comparison with earlier work,\cite{Coni83}  
where for $d=b=2$ and $q=1$ only
4 non-trivial fixed points were found. In the case of interest to us, 
the non-trivial fixed points are E,F,G,H and I which all have 
$\rho^* \simeq 1$.
(The other fixed points are trivial or merge into 
trivial ones for $q\rar 1$.)
Two of these (E and I) are independent of both
$q$ and $\eps$ and have one relevant eigenvalue. 
For the fixed point E, the relevant direction is characterized
by the exponent $1/\nu =d =1+\eps$, characteristic
of a (single) compact phase. For the fixed point I, 
the single relevant eigenvalue is
$1/\nu = d-1 < d$. We thus expect the fixed point E (I) to describe 
the compact (extended) phase of the lattice animal. 
The fixed point H has three relevant eigenvalues and therefore
represents the percolation fixed point 
(which is realized \cite{Giri77,Wu82} for $H=0$ and 
$L=0$ in eq.~(\ref{Potts})). 
F and G are tricritical points
and govern the renormalization group flow along the two critical lines 
leaving the percolation point H. We emphasise that the fixed point G
is only found when the limit $q\to 1^+$ is carefully taken. 

In summary, the collapse of an infinite branched polymer was studied using
its equivalence with the extended $q$--state Potts model in the $q\rar 1$ 
limit. The phase diagram was found from an exact 
solution on the Bethe lattice
and using a Migdal-Kadanoff renormalization group. In both cases
an extended and a {\em single} compact phase, 
separated by a line of $\theta$ transitions, was found. 
The $\theta$ line consists of two
segments which are in different universality classes and which meet at a 
multicritical point which coincides with the critical 
point of two-dimensional
percolation. This is in agreement with the 
available numerical results \cite{Fles92,Seno94,Pear95} in two dimensions. 
Our finding of a single compact phase
agrees with the transfer matrix results \cite{Seno94} but is in 
disagreement with the expectations based on graph 
enumeration studies.\cite{Fles92,Pear95}

\section*{Acknowledgments}
It is a pleasure to thank Flavio Seno and Julia Yeomans for the fruitful
collaboration which led to the work described here.

\end{document}